\documentclass[aps,prl,twocolumn,preprintnumbers,showpacs,nofootinbib]{revtex4}
\usepackage[english]{babel}
\usepackage{amsmath}
\usepackage{amssymb}
\usepackage{epsfig}
\usepackage{graphics,psfrag,rotating}
\usepackage{graphicx}
\usepackage{dcolumn}
\usepackage{bm}
\begin{document}
\newcommand{\Od}{{\cal O}}
\newcommand{\lsim}   {\mathrel{\mathop{\kern 0pt \rlap
  {\raise.2ex\hbox{$<$}}}
  \lower.9ex\hbox{\kern-.190em $\sim$}}}
\newcommand{\gsim}   {\mathrel{\mathop{\kern 0pt \rlap
  {\raise.2ex\hbox{$>$}}}
  \lower.9ex\hbox{\kern-.190em $\sim$}}}
\def\partials{{\partial \mkern-9mu/}}
\def\pslash{{p\mkern-7mu/}}

\title{The Higgs effective potential in the Littlest Higgs model
 at the one-loop level}

\author{Antonio Dobado, Lourdes Tabares-Cheluci}
\affiliation{Departamento de  F\'{\i}sica Te\'orica,
 Universidad Complutense de
  Madrid, 28040 Madrid, Spain}

\author{Siannah Pe{\~n}aranda}
\affiliation{IFIC - Instituto de F\'{\i}sica Corpuscular,
CSIC - Universitat de Val{\`e}ncia,\\ Apartado de Correos 22085,
E-46071 Valencia, Spain}
\date{\today}

\begin{abstract}
In this work we compute the contributions to the Higgs effective
potential coming from the fermion and gauge boson sectors at the
one-loop level in the context of the $SU(5)/SO(5)$ Littlest Higgs
(LH) model using a cutoff $\Lambda$ and including all finite
parts. We consider both, the $(SU(2)\times U(1))_{1} \times
(SU(2)\times U(1))_{2}$ and the $(SU(2)\times U(1))_{1} \times
(SU(2)\times U(1))$ gauge group versions of the LH model. We also
show that the Goldstone bosons present in the model do not
contribute to the effective potential at the one-loop level.
Finally, by neglecting the contribution of higher dimensional
operators, we discuss the restrictions that the new one-loop
contributions set on the parameter space of the LH model and the
need to include higher loop corrections to the Higgs potential.
\end{abstract}

\maketitle

\section{Introduction}

The quadratically divergent contributions to the Higgs mass and
the electroweak precision observables imply different scales for
physics beyond the Standard Model (SM), the first one below $2$
TeV and the second above $10$ TeV. This is the so called little
hierarchy problem. An interesting  attempt to solve it,  inspired
in an old suggestion by Georgi and Pais \cite {Georgi}, is the
Littlest Higgs model (LH) \cite{Cohen} which is based on a
$SU(5)/SO(5)$ non-sigma linear model (see \cite{Schmaltz} and
\cite{review1} for recent reviews).  Being a Goldstone boson (GB)
associated to this spontaneous symmetry breaking, the Higgs is
massless in principle. However one-loop corrections produce a
logarithmically divergent Higgs mass that could be compatible with
the present experimental bound of about $200$ GeV. The others GB
present in the model get quadratically divergent masses at the
one-loop level becoming very massive or give masses to the SM and
other additional gauge bosons present in the model through the
Higgs mechanism. All of these new states could give rise to a very
rich phenomenology that could be proved at the CERN Large Hadron
Collider (LHC) \cite{Logan}.

From the LH model it is possible, at least in principle, to compute
the Higgs low-energy effective potential. Obviously this effective
potential should reproduce the form of the SM potential, i.e.:
\begin{equation}\label{eq:Hpotential}
V_{eff}(H)=-\mu^{2} HH^{\dag}+\lambda (HH^{\dag})^{2},
\end{equation}
where $H=(H^0,H^{+})$ is the SM Higgs doublet and $\mu^2$ and
$\lambda$ are the well known Higgs mass and Higgs selfcoupling
parameters. Notice that, in order to have spontaneous symmetry
breaking of the electroweak symmetry, $\mu^2$ must be negative and
$\lambda$ must be positive to have a well defined energy minimum. In
addition these parameters should reproduce the SM relation $m_H^2= 2
\lambda v^2= 2\mu^2$ where $m_H$ is the Higgs mass and $v$ is the
vacuum expectation value (vev).

In principle $\mu^2$ and $\lambda$ receive contributions from
fermion, gauge boson and scalar loops, besides others that could
come from the ultraviolet completion of the LH model
\cite{italianos}. In this work we continue our program consisting in
the computation of the relevant terms of the Higgs low-energy
effective potential and their phenomenological consequences
including new restrictions on the parameter space of the LH model.

In particular we will study the consistency of the electroweak
symmetry breaking with the present experimental data in the case
in which, for the sake of simplicity, one neglects the
contribution of higher dimensional operators coming from the
ultraviolet completion of the LH model that are generically
present.

In \cite{ATP} we obtained the (one-loop) contribution coming from
the third generation quarks $t$ and $b$ plus the $T$ quark present
in the LH model. This contribution is essential since it provides
the right positive sign for $\mu^2$ being other contributions
negative.  Here we complete the one-loop computation of the Higgs
potential by including also gauge bosons and clarifying the role of
the GB at this level. We also discuss the validity of the one-loop
potential and the necessity of including some important higher loop
contributions to reproduce the expected value of the Higgs mass.

The paper is organized as follows: In Section 2 we briefly review
the LH model and set the notation. In section 3 we study the LH
model as a gauged non-linear sigma model (NLSM). In particular we
pay attention to the problem of the quartic divergencies appearing
when a cutoff $\Lambda$ is used to regulate the divergences of the
model and we show how they cancel at the one-loop level. We also
obtain the gauge fixing and Faddeev-Popov terms appropriate for the
calculation of the different gauge boson loops appearing latter in
our computations. In Section 4 we compute the effective potential at
the one-loop level.
 Section 5 is devoted to a discussion of our results and the
  constraints that our computation establishes on the LH
   parameters. Finally, in Section 6 we present the main conclusions
   of this paper and the prospects for future work.

\section{Setting of the Littlest Higgs model}

As it is well know the low energy dynamics of the LH model can be
described by a $(SU(2)\times U(1))_{1} \times (SU(2)\times
U(1))_{2}$ gauged non-linear
 sigma model based on the coset $K=G/H=SU(5)/SO(5)$. The Goldstone
 boson fields can be disposed  in a  $5 \times 5$ matrix $\Sigma$ given by:
 \begin{equation}
\Sigma=e^{2 i\Pi/f} \Sigma_{0},
\end{equation}
where:
\begin{equation}
\Sigma_{0}= \left(%
\begin{array}{ccc}
  0 & 0 & \textbf{1} \\
  0 & 1 & 0 \\
  \textbf{1} & 0 & 0 \\
\end{array}%
\right)
\end{equation}
has the proper $SU(5)$ symmetry breaking structure with
 $\textbf{1}$ being the $2 \times 2$ unit matrix, and
\begin{eqnarray}
\Pi & = & \left(%
\begin{array}{ccc}
\xi & \frac{-i}{\sqrt{2}}H^{\dag} & \phi^{\dag} \\
  \frac{i}{\sqrt{2}}H & 0 & \frac{-i}{\sqrt{2}}H^{*} \\
  \phi & \frac{i}{\sqrt{2}}H^{T} & \xi^{T} \\
\end{array}%
\right)        \\  \nonumber & + & \frac{1}{\sqrt{20}}\,\eta\,
{\mbox{diag }}(1,1,-4,1,1).\,,
\end{eqnarray}
Here  $H=(H^{0},H^{+})$ is the SM Higgs doublet, $\eta$ is the
real scalar and $\xi$ and $\phi$ are the real triplet and the
complex triplet respectively:
\begin{equation}
\xi=\left(%
\begin{array}{cc}
 \frac{1}{2} \xi^{0} & \frac{1}{\sqrt{2}}\xi^{+} \\
  \frac{1}{\sqrt{2}}\xi^{-} & -\frac{1}{2}\xi^{0}
\end{array}%
\right)\,,
\end {equation}
and
\begin{equation}
\phi=\left(%
\begin{array}{cc}
  \phi^{0} & \frac{1}{\sqrt{2}}\phi^{+} \\
\frac{1}{\sqrt{2}}\phi^{+} & \phi^{++}
\end{array}%
\right)\,.
\end{equation}
The gauged non-linear sigma model lagrangian describing the
low-energy GB and gauge boson dynamics is given by:
\begin{equation} \label{L0}
\textit{L}_{0} = \frac{f^2}{8}\mbox{tr} [(D_{\mu}\Sigma)
(D^{\mu}\Sigma)^\dag].
\end{equation}
The covariant derivative is defined as:
\begin{eqnarray}
D_{\mu}\Sigma & = &
\partial_{\mu}\Sigma-i\sum_{j=1}^{2}g_jW^a_j(Q_j^a\Sigma +\Sigma
Q_j^{aT})\nonumber\\
 & - & i\sum_{j=1}^{2}g'_jB_j(Y_j\Sigma+\Sigma Y_j^{T})
\end{eqnarray}
where $g$ and $g'$ are the gauge couplings,
 $Q_{1ij}^a=\sigma_{ij}^a/2$, for $i,j=1,2$, $Q_{2ij}^{a}=\sigma_{ij}^{a*}/2$ for
 $i,j=4,5$ and zero otherwise, $Y_{1}= diag(-3,-3,2,2,2)/10$ and
$Y_{2}= diag(-2,-2,-2,3,3)/10$. Diagonalizing the gauge boson mass
matrix  in this Lagrangian one realizes that the $W$ and $B$ SM
gauge bosons are massless and the $W'$ and $B'$ gauge bosons have
masses:
\begin{equation}
\label{eq:heavyGB} M_{W'}= f \sqrt{g_1^2+g_2^2}/2 {\mbox{ ; }}
M_{B'}= f \sqrt{g_1^{'2}+g_2^{'2}}/\sqrt{20}.
\end{equation}
The gauge bosons mass eigenstates are defined such as:
\begin{eqnarray}
W^a   & = & c_{\psi} W_1^a+ s_{\psi} W_2^a \nonumber\\
 W^{'a}  & = & s_{\psi} W_1^a- c_{\psi} W_2^a
\end{eqnarray}
where
\begin{eqnarray}
s_{\psi} &=& \sin \psi = \frac{g_1}{\sqrt{g_1^2+g_2^2}} \nonumber\\
 c_{\psi} &=& \cos \psi =
\frac{g_2}{\sqrt{g_1^2+g_2^2}}
\end{eqnarray}
and
\begin{eqnarray}
B  & = & c'_{\psi} B_1+ s'_{\psi} B_2 \nonumber\\
B'  & = & s'_{\psi} B_1- c'_{\psi} W_2
\end{eqnarray}
with
\begin{eqnarray}
s'_{\psi}=  \sin \psi' = \frac{g'_{1}}{\sqrt{{g'}_{1}^{\,2}+{g'}_{2}^{\,2}}}
\nonumber\\
 c'_{\psi}= \cos \psi' =
\frac{{g'}_{2}}{\sqrt{{g'}_{1}^{\,2}+{g'}_{2}^{\,2}}}\,.
\end{eqnarray}

A modified version of the LH model, such that the gauge subgroup
of $SU(5)$ is $[SU(2) \times SU(2)\times U(1)_Y]$ rather than
$[SU(2)\times U(1)_Y]^2$, has also been introduced~\cite{Peskin}.
In this case, the covariant derivative is defined as:
\begin{equation}
D_{\mu}\Sigma=\partial_{\mu}\Sigma
-i\sum_{j=1}^{2}g_jW^a_j(Q_j^a\Sigma +\Sigma Q_j^{aT})
-i g' B (Y \Sigma+\Sigma Y^{T})\,,
\end{equation}
where the generators $Q_{j}^{a}$ are the same as in the previous case,
and $Y=\frac{1}{2}$diag$(-1,-1,0,1,1)$. The field content of the matrix
$\Pi$ in $\Sigma$ is the same as in the LH model but there is no $B^{'}$
now. We consider in our analysis these two different models: the
original LH with two $U(1)$ groups ({\it{Model I}}) and the other
one with just one $U(1)$ group ({\it{Model II}}).

Then, at the tree level, the $SU(2)_L\times U(1)_Y$ SM gauge group
remains unbroken. The spontaneously symmetry breaking of this
group is expected to be produced in principle radiatively, mainly
due to the effect of the virtual quark fields from the third
generation, which give rise to an appropriate effective potential
for the SM Higgs doublet. These quarks will initially be denoted
by $u$ and $b$ and the additional vector-like quark will be
denoted by $U$. The interactions between these fermions and the
Goldstone bosons are given by the Yukawa Lagrangian:
\begin{equation}
\textit{L}_{YK}=-\frac{\lambda_{1}}{2}f
\overline{u}_{R}\epsilon_{mn}\epsilon_{ijk}\Sigma_{im}\Sigma_{jn}\chi_{Lk}-\lambda_{2}
f \overline{U}_{R}U_{L}+\mbox{h.c.}, \label{lagran}
\end{equation}
where $m,n=4,5$, $i,j=1,2,3$, and
\begin{eqnarray}
\overline{u}_{R}&=& c \,\overline{t}_{R}+ s  \,\overline{T}_{R}\,,\nonumber\\
\overline{U}_{R}&=&-s \,\overline{t}_{R}+   c \,\overline{T}_{R},
\end{eqnarray}
with:
\begin{eqnarray}
c&=&\cos
\theta=\frac{\lambda_{2}}{\sqrt{\lambda_{1}^{2}+\lambda_{2}^{2}}},\nonumber\\
s&=&\sin \theta =
\frac{\lambda_{1}}{\sqrt{\lambda_{1}^{2}+\lambda_{2}^{2}}}\,,
\end{eqnarray}
and
\begin{equation}
\chi_{L}=\left(%
\begin{array}{c}
  u \\
  b \\
  U \\
\end{array}%
\right)_{L}=\left(%
\begin{array}{c}
  t \\
  b \\
  T \\
\end{array}%
\right)_{L}.
\end{equation}
Here $b\,, t$ and $T$ are the mass eigenvectors coming from the
mass matrix included in the Yukawa Lagrangian with eigenvalues:
$m_{t}=m_b=0$ and $m_{T}=f\sqrt{\lambda_{1}^{2}+\lambda_{2}^{2}}$.
Notice that, contrary to the quark $T$ which is massive already at
this level, the  $t$ quark is massless and acquires mass only when
the electroweak symmetry is broken.

Then, the Yukawa Lagrangian can be written as:
\begin{equation}
\textit{L}_{\rm{Yuk}}=\overline{\chi}_{R} \hat{I}_{3x3}
\chi_{L}+\mbox{h.c.}\,,
\end{equation}
with
$$\chi_{R}=\left(%
\begin{array}{c}
  t \\
  b \\
  T \\
\end{array}%
\right)_{R}$$ and $\hat{I}_{3x3}$ the Higgs-quark interaction
matrix is given by
\begin{eqnarray}
\hat{I}=\left(%
\begin{array}{ccc}
 -\sqrt{2}\lambda_{1}cH^{0}\Theta
 &-\sqrt{2}\lambda_{1}cH^{+}
 \Theta &\lambda_{1}c \frac{HH^{\dag}}{f}\Theta^{'}\\
0 & 0 & 0 \\
-\sqrt{2}\lambda_{1}sH^{0}\Theta & -\sqrt{2}\lambda_{1}sH^{+}
\Theta &\lambda_{1}s \frac{HH^{\dag}}{f} \Theta^{'}
\end{array}%
\right),
\end{eqnarray}
where $\Theta$ and $\Theta^{'}$ are functions on $H H^\dag/ f^2$
whose expansion starts as:
\begin{eqnarray}
\Theta\left(\frac{HH^{\dag}}{f^2}\right)=1-\frac{2 HH^{\dag}}{3 f^{2}}+...\\
\nonumber
\Theta'\left(\frac{HH^{\dag}}{f^2}\right)=1-\frac{HH^{\dag}}{3f^{2}}+...
\end{eqnarray}
Thus the complete Lagrangian for the quarks is:
\begin{eqnarray}
\textit{L}_{\chi}&=&\textit{L}_{0}+\textit{L}_{\rm{Yuk}}=
\overline{\chi}_{R}(i
\partials-M+\hat{I})
\chi_{L}+\mbox{h.c.}
\end{eqnarray}with $M=$diag$(0,0,m_T)$.

Since we are interested in the computation of the
contribution to the SM Higgs $H$ effective potential, we
can set $\xi= \phi = \eta = 0 $.

\section{The Littlest Higgs model as a Non linear sigma model}

\subsection{Quartic divergences}

In order to compute the contributions to the Higgs potential
coming from scalar and gauge boson loops it is useful to study the
LH model as a particular case of gauged non-linear sigma model
(NLSM) based on the coset $K=G/H=SU(5)/SO(5)$ (see  ~\cite{book}
for a review on gauged NLSM). To start with we will turn off the
gauge interactions by taken $g_{i}^{(')}=0$. Then the
$\textit{L}_0$ lagrangian is:
\begin{equation}
\textit{L}_0=\frac{f^{2}}{8}\mbox{tr}[(\partial_{\mu}\Sigma)(\partial^{\mu}\Sigma)^{\dag}]
\end{equation}
This lagrangian can be written also as a NLSM lagrangian
\begin{equation}
\textit{L}_{0}=\frac{1}{2} g_{\alpha
\beta}(\pi)\partial_{\mu}\pi^{\alpha}\partial^{\mu}\pi^{\beta}
\end{equation}
where $\pi^{\alpha}$ are Gaussian coordinates on $K$ and the $K$
metric is defined as:
\begin{equation}
g_{\alpha \beta}\equiv\frac{f^{2}}{4}\mbox{tr}\frac{\partial
 \Sigma}{\partial \pi^{\alpha}}\frac{\partial \Sigma^{\dag}}{\partial \pi^{\beta}}
\end{equation}
This metric can be split as
\begin{equation}
g_{\alpha \beta}=\delta_{\alpha \beta}+\Delta_{\alpha \beta}(\pi)
\end{equation}
where
\begin{eqnarray}
\Delta_{ab}(\pi)&=&-\frac{8}{3!f^{2}} \mbox{tr}(T_{\alpha}
T_{\beta}T_{\delta}T_{\gamma}+T_{\alpha}T_{\delta}T_{\beta}
T_{\gamma}\nonumber \\ &+&T_{\alpha}T_{\delta}T_{\gamma}
T_{\beta})\pi_{\delta}\pi_{\gamma} + \textit{O}(\pi^{4}) \nonumber
\\ &=&-\frac{8}{3!f^{2}}\kappa_{\alpha \beta \delta
\gamma}\pi_{\delta}\pi_{\gamma}+ \textit{O}(\pi^{4})
\end{eqnarray}
and we have written $\Pi$ as $\Pi=\pi^{\alpha}T^{\alpha}$ with the
$T^{\alpha}$ matrices normalized so that
tr$T^{\alpha}T^{\beta}=\delta^{\alpha\beta}$. Now we consider the
coupling of the NLSM with any other field $\phi$ which for
simplicity will be taken to be a real scalar. The corresponding
action can be written as $S[\pi,\phi]=S_{0}[\pi]+S'[\pi,\phi]$. The
effective action for the $\phi$ field can be obtained by integrating
out the GB fields $\pi$. However this integration is not trivial at
all. Due to the geometrical nature of the NLSM, not only its action,
but also the integration measure, must be $G$ invariant and
covariant in the $K$ coset sense \cite{measure}. Thus the proper
$\phi$ effective action is given by:
\begin{equation}
e^{i \Gamma[\phi]}=\int [d\pi \sqrt{g}]e^{iS[\pi,\phi]}
\end{equation}
The measure factor $\sqrt{g}$ can now be exponentiated to find
\begin{equation}
e^{i \Gamma[\phi]}=\int [d\pi]e^{i(S[\pi,\phi]+\Delta\Gamma[\pi])}
\end{equation}
with
\begin{eqnarray}
\Delta \Gamma[\pi]&=&-\frac{i}{2}\delta(0)\int dx \log(1+\Delta)\nonumber \\
&=&-\frac{i}{2}\delta(0)\Sigma_{k=1}^{\infty}\frac{(-1)^{k+1}}{k}\int dx \mbox{tr}\Delta^{k}
\end{eqnarray}
where by using the notation $d\tilde k \equiv d^D k/(2 \pi)^D$
with $D$ being the space-time dimensionality so that:
\begin{equation}
\delta(0)=\int d\tilde{k}
\end{equation}
In the dimensional regularization scheme one has
\begin{equation}
\delta(0)=\int \frac{d^{D}k}{(2\pi)^{D}}=0
\end{equation}
but using an ultraviolet cutoff $\Lambda$ to define divergent
integrals
\begin{equation}
\delta(0)=\int d\tilde{k}=i\frac{\Lambda^{4}}{2(4\pi)^{2}}
\end{equation}
which obviously does not vanish. Then, in order to take into
account the invariant measure effects in the NLSM one needs to add
to the classical lagrangian the term
\begin{equation}
S_{0} \rightarrow S_{0}+\Delta\Gamma
\end{equation}
whenever one is not using dimensional regularization. It is not
difficult to see that this term is formally of the same order as
the one-loop contributions.

On the other hand the GB contribution to the Higgs effective
potential is defined as
\begin{equation}
\Gamma_{eff}[\overline{\pi}]=-\int dx V_{eff}(\overline{\pi})
\end{equation}
where $\overline{\pi}$ is a constant field and
\begin{equation}
e^{i \Gamma_{eff}[\overline{\pi}]}=\int
[d\pi'\sqrt{g}]e^{iS_{0}[\overline{\pi}+\pi']}
\end{equation}
with
\begin{equation}
\frac{\delta \Gamma_{eff}[\pi]}{\delta \pi}|_{\pi=\overline{\pi}}=0
\end{equation}
At the one-loop level the last equation can be written as
\begin{equation}
\frac{\delta S_{0}[\pi]}{\delta \pi}|_{\pi=\overline{\pi}} \simeq
0
\end{equation}
and then the NLSM action can be expanded as
\begin{eqnarray}
S_{0}[\overline{\pi}+\pi']&=&S_{0}[\overline{\pi}]\nonumber \\
&+&\frac{1}{2}\int dx dy
\pi'^{\alpha}(x)\frac{\delta^{2}S_{0}}{\delta \pi^{\alpha}
(x)\delta\pi^{\beta}(y)}\Big\vert_{\pi=\overline{\pi}} \pi'^{\beta}(y) \nonumber \\
\end{eqnarray}
Therefore we have
\begin{equation}
\Gamma_{eff}[\overline{\pi}]=S_{0}[\overline{\pi}]+\frac{i}{2}\mbox{Tr
log}(1+GO)+...
\end{equation}
where the inverse GB propagator is
\begin{equation}
(G_{xy}^{\alpha \beta})^{-1}=- \Box_{x}\delta_{xy}\delta_{\alpha
\beta}
\end{equation}
$\delta_{xy}$ being the short for $\delta(x-y)$ and
\begin{equation}
O_{xy}^{\alpha \beta} = \frac{\delta^{2}}{\delta
\pi^{\alpha}(x)\delta \pi^{\beta}(y)}\int
 dx \partial_{\mu}\pi^{\alpha}\Delta_{\alpha \beta}(\pi)\partial^{\mu}\pi^{\beta}\Big\vert_{\pi=\overline{\pi}}
\end{equation}
In order to compute the Higgs effective action we only need to
consider the case $ \overline{\pi}=$cte which means
$\partial_{\mu}\overline{\pi}=0$ and then we have
\begin{equation}
O_{xy}^{\alpha \beta}(\pi)=-\Delta_{\alpha
\beta}(\overline{\pi})\Box_{x}\delta_{xy}.
\end{equation}
Therefore we get
\begin{equation}
\Gamma_{eff}[\overline{\pi}]=\frac{i}{2}
\mbox{Tr}\Sigma_{k=1}^{\infty}\frac{(-1)^{k+1}}{k}(OG)^{k}+...
\end{equation}
or
\begin{equation}
\Gamma_{eff}[\overline{\pi}]=\frac{i}{2}\delta(0)\int dx
\mbox{tr}\left(\Delta-\frac{\Delta^{2}}{2}+\frac{\Delta^{3}}{3}+...\right)+...
\end{equation}
This effective action has exactly the same form as the measure
term discussed above so finally we get:
\begin{equation}
\Gamma_{eff}[\overline{\pi}]+\Delta\Gamma[\overline{\pi}]=0
\end{equation}
Therefore we arrive to the important conclusion that the GB do
not contribute to the Higgs potential in any NLSM at the one loop
level and in particular this is the case for the $SU(5)/SO(5)$ LH
model.

\subsection{Gauge fixing and the Faddeev-Popov terms}

In the following we will concentrate on the gauge bosons in order
to be able to compute their contribution to the Higgs effective
potential. Thus we turn on again the gauge boson fields in the
NLSM:
\begin{equation}
\textit{L}_{0}=\frac{f^{2}}{8}\mbox{tr}[(D_{\mu}\Sigma)(D_{\mu}\Sigma)^{\dag}]
\end{equation}
The covariant derivative can be written in terms of the mass
eigenstates as:
\begin{eqnarray}
D_{\mu}\Sigma&=&\partial_{\mu}\Sigma-i g
W^{a}_{\mu}(Q^{a}_{L}\Sigma+ \Sigma
Q_{L}^{aT})\nonumber \\&-&i g_{R}
W'^{a}_{\mu}(Q^{a}_{R}\Sigma
+ \Sigma Q_{R}^{aT})
-ig'B_{\mu}(Y\Sigma+\Sigma Y^{T}))\nonumber \\
&-&ig''B'_{\mu}(Y'\Sigma+\Sigma Y'^{T}))
\end{eqnarray}
where the different couplings and generators are defined as:
\begin{eqnarray}
\label{eq:extradef}
Q_{L}^{a} &\equiv& Q_{1}^{a}+Q_{2}^{a},      \nonumber \\
g &\equiv& g_{L}=\frac{g_{1}g_{2}}{\sqrt{g_{1}^{2}+g_{2}^{2}}}, \nonumber \\
Y &\equiv& Y_{1}+Y_{2},    \nonumber \\
g'&=&\frac{g'_{1}g'_{2}}{\sqrt{g_{1}^{'2}+g_{2}^{'2}}}, \nonumber \\
g_{R}Q_{R}^{a} &\equiv& \frac{g_{1}^{2}Q_{1}^{a}-g_{2}^{2}Q_{2}^{a}}
{\sqrt{g_{1}^{2}+g_{2}^{2}}}, \nonumber \\
g_{R}^{2} &\equiv& \frac{1}{2}(g_{1}^{2}s_{\psi}^{2}+g_{2}^{2}c_{\psi}^{2}), \nonumber \\
g''Y' &\equiv& \frac{g_{1}^{'2}Y_{1}-g_{2}^{'2}Y_{2}}{\sqrt{g_{1}^{'2}+g_{2}^{'2}}},
 \nonumber \\
g''^{2} &\equiv&
\frac{1}{2}(g_{1}^{'2}s_{\psi'}^{2}+g_{2}^{'2}c_{\psi'}^{2}).
\end{eqnarray}
The first four definitions correspond to the diagonal group
$(SU(2) \times U(1))_{1+2}$ and
 the last four to the axial group $(SU(2)\times U(1))_{1-2}$.
Notice that $g_{R}$ and $g''$ are functions of the mixing angles
$\psi$ (for the $SU(2)$ group) and $\psi'$ (for the $U(1)$ group).

Expanding the Lagrangian we obtain the gauge and GB mixed terms:
\begin{eqnarray} \label{mixterm}
\textit{L}_{W \Pi}&=&\textit{L}_{B \Pi}=0, \nonumber\\
\textit{L}_{W' \Pi}&=&M'_{W}\partial^{\mu}W_{\mu}^{'a} \xi^{a},\nonumber\\
\textit{L}_{B' \Pi}&=&M'_{B}\partial^{\mu}B'_{\mu} \eta,
\end{eqnarray}
where $\xi^{1}=(\xi^{+}+\xi^{-})/\sqrt{2}$,
$\xi^{2}=i(\xi^{+}-\xi^{-})/\sqrt{2}$, $\xi^{0}=\xi^{3}$ and $\eta$
are the GB which will give masses to the heavy gauge bosons $W'$ and
$B'$. Following the standard Faddeev-Popov procedure it is not
difficult to find the gauge fixing and the ghost Lagrangian. The
first one is given by
\begin{eqnarray}
\textit{L}_{GF}&=&-\frac{1}{2 \alpha'}(\partial^{\mu}W'^{a}_{\mu}+
\alpha' M'_{W}\xi^{a})^{2}\nonumber \\
&-&\frac{1}{2 \beta'}(\partial^{\mu}B'_{\mu}+\beta' M'_{B}\eta)^{2}\nonumber \\
&-&\frac{1}{2 \alpha}(\partial^{\mu}W^{a}_{\mu})^{2}-
\frac{1}{2 \beta}(\partial^{\mu}B_{\mu})^{2},
\end{eqnarray}
which cancels the unwanted mixing terms (\ref{mixterm}) and makes
the propagator well defined. For a gauge boson $A_{\mu}^{a}$ the
Faddeev-Popov Lagrangian is
\begin{equation}
\textit{L}_{FP}=\int dy \overline{c}^{a}(x)\frac{\delta f^{a}(x)}{\delta \theta^{b}(y)}c^{b}(y)
\end{equation}
where
\begin{equation}
f^{a}(A_{\mu}^{a},\pi^{a}) = \partial^{\mu}A_{\mu}^{a}+\alpha
M_{A}\pi^{a}.
\end{equation}
In the general case the effect of the gauge transformations on the
GB and the gauge boson fields $A_{\mu}^{a}$ will be
\begin{equation}
\pi'^{\alpha}=\pi^{\alpha}+\kappa^{\alpha}_a(\pi)\theta^{a}(x)
\end{equation}
and
\begin{equation}
A'^{a}_{\mu}=A^{a}_{\mu}-\partial_{\mu}\theta^{a}+gC_{abc}\theta^{b}A_{\mu}^{c}
\end{equation}
where $\kappa^{\alpha}_a(\pi)$ are the Killing vectors corresponding
to the gauge symmetry on the coset and $C_{abc}$ are the structure
constants. The covariant derivative is defined as:
\begin{equation}
D_{\mu}\pi^{\alpha}=\partial_{\mu}\pi^{\alpha}-g\kappa^{\alpha}_{a}(\pi)A_{\mu}^{a}
\end{equation}
so that the gauge fixing terms is:
\begin{equation}
f^{a}(A_{\mu}^{a},\pi^{a}) = \partial^{\mu}A_{\mu}^{a}+g\alpha
\kappa_{\alpha}^a \pi^{\alpha}
\end{equation}
and the Faddeev-Popov Lagrangian can be written as:
\begin{equation}
\textit{L}_{FP}=\overline{c}_{a} \Box c_{a}
-gC_{abc}\overline{c}^{a}\partial_{\mu}c^{b}A^{\mu c}+ g \alpha
\kappa^{a}_{\alpha}\kappa^{\alpha b}\overline{c}_{a}c_{b}.
\end{equation}
Then ghost-GB interaction is given by
\begin{equation}
\Delta \textit{L}=g \alpha \kappa^{a}_{\alpha}\kappa^{\alpha
b}\overline{c}_{a}c_{b}.
\end{equation}
Therefore  if we work in the Landau gauge, that is $\alpha=0$,
there are no ghost-GB interactions. This fact will be useful later
for the computation of the gauge boson contribution to the Higgs
effective potential.

In this gauge the quadratic part of the gauge boson lagrangian is
just:
\begin{equation}
\textit{L}_{\Omega}=\frac{1}{2}\Omega^{\mu}((\Box+M_{\Omega}^{2})g_{\mu\nu}
-\partial_{\mu}\partial_{\nu}+2 \tilde{I} \,g_{\mu\nu})\Omega^{\nu}
\end{equation}
where $\Omega$ stands for any of the gauge bosons:
\begin{equation}
\Omega^{\mu}=(W'^{\mu a},W^{\mu a},B'^{\mu},B^{\mu}),\\
\end{equation}
which are the mass matrix eigenstates with masses
\begin{equation}
M_{\Omega}=(M_{W'} 1_{3\times 3},0_{3\times\ 3},M_{B'},0),
\end{equation}
and $\tilde{I}_{6\times 6}$ is the interaction matrix between the
gauge bosons and the Higgs doublet, given in Appendix A.

\section{The effective potential}

In order to obtain the one-loop Higgs effective potential we
consider constant Higgs fields, i.e. $\partial_{\mu}H=0$. Thus we
have:
\begin{equation}
S_{eff}[\overline{H}]=-\int dx V_{eff}(\overline{H}).
\end{equation}
Remember that we found that GB do not contribute to this effective
potential at the one-loop level. In addition, by using the Landau
gauge, we do not have to consider any ghost field at this level.
Then the effective action is obtained just by integrating out the
$b$, $t$ and $T$ fermions and the $W$, $W'$, $B$ and $B'$ gauge
bosons:
\begin{equation}
e^{iS_{eff}[H]}=\int [d\chi
d\overline{\chi}][d\Omega]e^{iS[H,\chi,\Omega]},
\end{equation}
with
\begin{equation}
S[H,\chi,\Omega]=\int
dx(\partial_{\mu}H\partial^{\mu}H+\textit{L}_{\chi}+\textit{L}_{\Omega}),
\end{equation}
By using standard techniques (see for instance ~\cite{book}) the
one-loop effective action can be written as:
\begin{equation}
\label{eq:action} S_{eff}[H]=\int
dx(\partial_{\mu}H\partial^{\mu}H+S_{f}[H]+S_{g}[H])
\end{equation}
And  then, the effective potential can be written as:
\begin{equation}
V_{eff}(H)=V_{f}(H)+V_{g}(H)
\end{equation}
where $V_{f}(H)$ and $V_{g}(H)$ are the fermionic and gauge boson
contribution to the Higgs effective potential respectively. The
general form of the effective potential is
\begin{eqnarray}
V_{eff}(H)=-\mu^{2}HH^{\dag}+\lambda (HH^{\dag})^{2}+...
\end{eqnarray}
where we keep only the first two terms which are the relevant ones
for the electroweak symmetry breaking and, in particular, for the
computation of the Higgs mass. Obviously, at the one-loop level
the $\mu^2$ and the $\lambda$ parameters have separated
contributions from the fermionic and the gauge sector,
\begin{eqnarray}
\mu^{2} & = & \mu^{2}_f+\mu^{2}_g \,,  \nonumber \\
\lambda & = & \lambda_f +  \lambda_g\,.
\end{eqnarray}

\subsection{Fermionic contribution}

In this case the one-loop computation is exact since the action is
quadratic in the fermionic fields corresponding to the $t$, $b$
and $T$ quarks. Details of the computation of the fermionic
contribution to the effective action and the Higgs effective
potential parameters, $\mu^2_f$ and $\lambda_f$, are given
in~\cite{ATP} (see Fig.1 for the contributing Feynman diagrams).
For the purpose of illustration and the final discussion of this
paper, we summarize here the fermionic contribution at one loop
level to these Higgs potential parameters:
\begin{equation}
\mu^{2}_f= N_{c} \frac{m_{T}^{2} \lambda_{t}^{2}}{4 \pi^{2}}
\log\left(1+\frac{\Lambda^{2}}{m_{T}^{2}}\right),
\label{eq:muparameter}
\end{equation}
and
\begin{eqnarray}
\hspace{-0.5cm}\lambda_f&=&\frac{N_{c}}{(4
\pi)^{2}}\left[2(\lambda_{t}^{2}+\lambda_{T}^{2})
\frac{\Lambda^{2}}{f^{2}}\right.\nonumber\\
&-&\left.\log\left(1+\frac{\Lambda^{2}}{m_{T}^{2}}\right)
\left(-\frac{2m_{T}^{2}}{f^{2}}\left(\frac{5}{3}
\lambda_{t}^{2}+\lambda_T^{2}\right)+4\lambda_{t}^{4}\right.\right.\nonumber\\
&+&4\left.\left.(\lambda_{T}^{2}+
\lambda_{t}^{2})^{2}\right)\right.\nonumber\\
&-&4\lambda_{T}^{2}\frac{1}{1+\frac{m_{T}^{2}}{\Lambda^{2}}}\left
(\frac{m_{T}^{2}}{f^{2}}-2\lambda_{t}^{2} -\lambda_{T}^{2}\right)\nonumber\\
&-&\left.4\lambda_{t}^{4}
\log\left(\frac{\Lambda^{2}}{m^{2}}\right)\right]\,,
 \label{eq:lambdaTOP}
\end{eqnarray}
where $N_c$ is the number of colors and,  $\lambda_t$ and
$\lambda_T$ are, respectively, the SM top Yukawa coupling and the
heavy top Yukawa coupling, given by:
\begin{equation}
\lambda_{t}=
\frac{\lambda_{1}\lambda_{2}}{\sqrt{\lambda_{1}^{2}+\lambda_{2}^{2}}}\,,\,\,\,\,\,\,
\lambda_{T}=\frac{\lambda_{1}^{2}}{\sqrt{\lambda_{1}^{2}+\lambda_{2}^{2}}}\,.
\end{equation}

\begin{figure}[tbh!]
\begin{center}
\epsfig{file=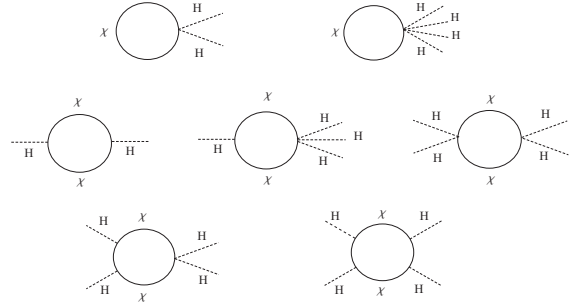,scale=0.40} \caption{Fermionic one-loop
diagrams contributing to the Higgs potential with $\chi = t, b$ or
$T$. All possible combinations of these particles appear in the
loops.} \label{diagrams1}
\end{center}
\end{figure}

\subsection{Gauge bosons contribution}

Here we concentrate in the gauge boson contribution at one loop
level to the Higgs effective action, $S_{g}[H]$. We use the Landau
gauge for the reasons discussed above. The Higgs effective action
can be expanded as:
\begin{equation}\label{exp}
S_{g}[\overline{H}]=\frac{i}{2}\mbox{Tr}\log[1+2G\tilde{I}]=\frac{i}{2}\Sigma_{k=1}
^{\infty}\frac{(-1)^{k+1}}{k}\mbox{Tr}(2 G\tilde{I})^{k}
\end{equation}
where the gauge boson propagators are given by:
\begin{equation}
G_{ab}^{\Omega\mu\nu}(x,y)=\int d\tilde{k}\frac{e^{-ik(x-y)}}{k^{2}-M_{\Omega}^{2}}
\left(-g^{\mu\nu}+\frac{k^{\mu}k^{\nu}}{k^{2}-M_{\Omega}^{2}}\right)_{ab}
\end{equation}
and the interaction operators are:
\begin{equation}
\hat{\tilde{I}}^{ab}(x,y)=(\tilde{I}_{2}(x)+\tilde{I}_{4}(x))\delta(x-y)\delta^{ab}
\end{equation}

\begin{figure}[tbh!]
\begin{center}
\epsfig{file=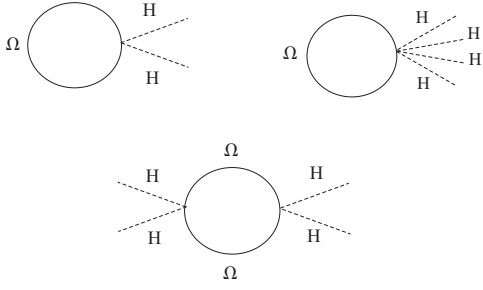,scale=0.60} \caption{Gauge bosons loops
contributing to the Higgs effective potential. Here $\Omega =
W^{'1,2,3}, W^{1,2,3}, B'$ or $B$ and all possible combinations of
these boson appear in the loops.} \label{diagrams2}
\end{center}
\end{figure}
In order to obtain the gauge boson contribution to the $\mu$ and
$\lambda$ parameters we only need to consider the terms $k=1$ and
$k=2$ in the expansion (\ref{exp}). The generic one-loop diagrams
are shown in Fig.~\ref{diagrams2}.
Then we have to compute for $k=1$:
\begin{eqnarray}
S_{g}^{(1)}[\overline{H}]&=&i\mbox{Tr}G(\tilde{I}_{2}+\tilde{I}_{4})\nonumber \\
&=&-i \Sigma_{a}\int dx\int d\tilde{k} \nonumber \\
&&\frac{1}{k^{2}-M_{a}^{2}}\left(g^{\mu\nu}-\frac{k^{\mu}k^{\nu}}{k^{2}}\right)
(\Delta_{2}^{aa}+\Delta_{4}^{aa})g_{\nu \mu}\nonumber \\
&=&-3\Sigma_{a}\int dx (\tilde{I}_{2}^{aa}+\tilde{I}_{4}^{aa}) I_{0}(M_{\Omega a})\,,
\end{eqnarray}
and for $k=2$:
\begin{eqnarray}
S_{g}^{(2)}[\overline{H}]&=&-i\left[G\tilde{I}_{2}G\tilde{I}_{2}\right]\nonumber \\
&=&-i \Sigma_{ab}\int dx dy d\tilde{k}d\tilde{p}\nonumber \\
&&\left[\left(g^{\mu\nu}-\frac{k^{\mu}k{\nu}}{k^{2}}\right)\frac{e^{-ik(x-y)}}
{k^{2}-M_{a}^{2}}(\tilde{I}_{2}^{ab}+\tilde{I}_{4}^{ab})\right.\nonumber \\
&&\left.\left(g_{\nu\mu}-\frac{p_{\nu}p_{\mu}}{k^{2}}\right)\frac{e^{-ip(y-x)}}
{p^{2}-M_{b}^{2}}(\tilde{I}_{2}^{ba}+\tilde{I}_{4}^{ba})\right]\nonumber \\
&=&-3 \Sigma_{ab}\int dx \tilde{I}_{2}^{ab}\tilde{I}_{2}^{ba}I(M_{a}^{2},M_{b}^{2})+
\textit{O}(H^{6})\,.\nonumber \\
\end{eqnarray}

After some work, these two terms are found to be:
\begin{widetext}
\begin{eqnarray}
S_{g}^{(1)}[\overline{H}]&=&-\frac{3}{(4\pi)^{2}}\int dx\left\{(\overline{H}
\overline{H}^{\dag})\left[\frac{3}{4}g^{2}M_{W'}^{2} \log\left(1+
\frac{\Lambda^{2}}{M_{W'}^{2}}\right)+\frac{1}{4}g'^{2}M_{B'}^{2}
\log\left(1+\frac{\Lambda^{2}}{M_{B'}^{2}}\right)\right]\right. \nonumber \\
&+&\left.(\overline{H}\overline{H}^{\dag})^{2}\left[\frac{(g_{1}^{2}+g_{2}^{2})
\Lambda^{2}}{16 f^{2}}+\frac{(g_{1}^{'2}+g_{2}^{'2})\Lambda^{2}}{16 f^{2}}-
\left(\frac{g^{2}}{4 f^{2}}+\frac{g_{1}^{2}+g_{2}^{2}}{16 f^{2}}\right)M_{W'}^{2}
\log\left(1+\frac{\Lambda^{2}}{M_{W'}^{2}}\right)\right.\right.\nonumber\\
&-&\left.\left.\left(\frac{g^{'2}}{12
f^{2}}+\frac{g_{1}^{'2}+g_{2}^{'2}}{16 f^{2}}
\right)M_{B'}^{2}\log\left(1+\frac{\Lambda^{2}}{M_{B'}^{2}}\right)
\right]\right\}
\end{eqnarray}
\end{widetext}
and
\begin{widetext}
\begin{eqnarray}
S_{g}^{(2)}[\overline{H}]&=& \frac{3}{(4\pi)^{2}} \int dx
(\overline{H}\overline{H}^{\dag})^{2}\left[\frac{3}{16} g^{4}
\left(\frac{-1}{1+\frac{M_{W'}^{2}}{\Lambda^{2}}}+\log
\left(1+\frac{\Lambda^{2}}{M_{W'}^{2}}\right)\right)
+g'^{4}\left(\frac{-1}{1+\frac{M_{B'}^{2}}{\Lambda^{2}}}+
\log\left(1+\frac{\Lambda^{2}}{M_{B'}^{2}}\right)\right)\right.\nonumber \\
&+&\left. 2\frac{(g_{1}^{2}g_{2}^{'2}+g_{2}^{2}g_{1}^{'2})^{2}}
{16(g_{1}^{2}+g_{2}^{2})(g_{1}^{'2}+g_{2}^{'2})}\frac{1}
{M_{W'}^{2}-M_{B'}^{2}} \times \left(M_{W'}^{2}\log
\left(1+\frac{\Lambda^{2}}{M_{W'}^{2}}\right)-M_{B'}^{2}\log
\left(1+\frac{\Lambda^{2}}{M_{B'}^{2}}\right)\right)\right.\nonumber \\
&+&(3 g^{2}+g^{'2})\frac{(g_{1}^{2}-g_{2}^{2})^{2}}{8 (g_{1}^{2}+g_{2}^{2})}
\log\left(1+\frac{\Lambda^{2}}{M_{W'}^{2}}\right)+(g^{2}+g^{'2}) \frac{(g_{1}^{'2}
-g_{2}^{'2})^{2}}{8 (g_{1}^{'2}+g_{2}^{'2})}\log\left(1+\frac{\Lambda^{2}}{M_{B'}^{2}}\right)\nonumber \\
&+&\left. (\frac{3}{16}g^{4}+ \frac{1}{16}g^{'4}+\frac{1}{2}g^{2}g^{'2})
\log\left(\frac{\Lambda^{2}}{m^{2}}\right)\right]\,.
\end{eqnarray}
\end{widetext}
From these effective actions we find, for the \emph{Model I},
\begin{eqnarray}
\label{eq:muM1}
\mu^{2}_{g}&=&-\frac{3}{64\pi^{2}}\left[3g^{2}M_{W'}^{2}
\log\left(1+\frac{\Lambda^{2}}{M_{W'}^{2}}\right)\right.\nonumber\\
&+&\left.g^{'2}M_{B'}^{2}\log\left(1+\frac{\Lambda^{2}}{M_{B'}^{2}}
\right)\right]\,,
\end{eqnarray}
\begin{widetext}
\begin{eqnarray}
\label{eq:lambdaM1}
\lambda_{g}&=&-\frac{3}{(16\pi f)^{2}}\left[-\left(\frac{g^{2}}{c_{\psi}^{2}s_{\psi}^{2}}+
\frac{g^{'2}}{c_{\psi}^{'2}s_{\psi}^{'2}}\right)\Lambda^{2}+g^{2}M_{W'}^{2}
\log\left(1+\frac{\Lambda^{2}}{M_{W'}^{2}}\right)\left(4+\frac{1}{c_{\psi}^{2}s_{\psi}^{2}}
+2g^{'2}\frac{(c_{\psi}^{2}s_{\psi}^{'2}+s_{\psi}^{2}c_{\psi}^{'2})^{2}}{c_{\psi}^{2}
s_{\psi}^{2}c_{\psi}^{'2}s_{\psi}^{'2}}\frac{f^{2}}{M_{W'}^{2}-M_{B'}^{2}}\right)
\right.\nonumber \\
&+&\left.g^{'2}M_{B'}^{2} \log\left(1+\frac{\Lambda^{2}}{M_{B'}^{2}}\right)
\left(\frac{4}{3}+\frac{1}{c_{\psi}^{'2}s_{\psi}^{'2}}
+2g^{2}\frac{(c_{\psi}^{2}s_{\psi}^{'2}+s_{\psi}^{2}c_{\psi}^{'2})^{2}}{c_{\psi}^{2}
s_{\psi}^{2}c_{\psi}^{'2}s_{\psi}^{'2}}\frac{f^{2}}{M_{B'}^{2}-M_{W'}^{2}}\right)
\right.\nonumber \\
&+&\left.f^{2}\log\left(1+\frac{\Lambda^{2}}{M_{W'}^{2}}\right)\left(3g^{4}+
2(3g^{2}+g^{'2})g^{2}\frac{(s_{\psi}^{2}-c_{\psi}^{2})^{2}}{c_{\psi}^{2}s_{\psi}^{2}}\right)
+f^{2}\log\left(1+\frac{\Lambda^{2}}{M_{B'}^{2}}\right)\left(g^{'4}+2(g^{2}+
g^{'2})g^{'2}\frac{(s_{\psi}^{'2}-c_{\psi}^{'2})^{2}}{c_{\psi}^{'2}s_{\psi}^{'2}}
\right)\right.\nonumber \\
&+&\left.f^{2}\log\left(\frac{\Lambda^{2}}{m^{2}}\right)\left(3g^{4}+g^{'4}+
8g^{2}g^{'2}\right)-3f^{2}\frac{g^{4}}{1-\frac{M_{W'}^{2}}{\Lambda^{2}}}-
f^{2}\frac{g^{'4}}{1-\frac{M_{B'}^{2}}{\Lambda^{2}}}\right]\,.
\end{eqnarray}
\end{widetext}
In the context of \emph{Model II}, a similar computation gives:
\begin{equation}
\label{eq:muM2}
\mu_g^{2}=-\frac{3}{64\pi^{2}}\left(3g^{2}M_{W'}^{2}
\log\left(1+\frac{\Lambda^{2}}{M_{W'}^{2}}\right)+g^{'2} \Lambda^{2}\right),
\end{equation}
\begin{widetext}
\begin{eqnarray}
\label{eq:lambdaM2} \lambda_g&=&-\frac{3}{(16 \pi f)^{2}}
\left[-\frac{g^{2}}{c_{\psi}^{2}s_{\psi}^{2}}\Lambda^{2}+
\frac{16}{12}g'^{2}\Lambda^{2}+ g^{2}M_{W'}^{2}
\log\left(\frac{\Lambda^{2}}{M_{W'}^{2}}+1\right)
\left(4+\frac{1}{c_{\psi}^{2}s_{\psi}^{2}}\right) \right. \nonumber \\
&+&\left.f^{2}\log \left(1+\frac{\Lambda^{2}}{M_{W'}^{2}}\right)
\left(3g^{4}+2(3g^{2}+g'^{2})g^{2}\frac{(s_{\psi}^{2}-c_{\psi}^{2})^{2}}
{s_{\psi}^{2}c_{\psi}^{2}}\right)\right. \nonumber\\
&+&\left.f^{2}\log\left(\frac{\Lambda^{2}}{m^{2}}\right)
(3g^{4}+g'^{4}+8g^{2}g'^{2})
-3f^{2}\frac{g^{4}}{1-\frac{M_{W'}^{2}}{\Lambda^{2}}}\right].
\end{eqnarray}
\end{widetext}
To summarize the fermion and gauge boson contribution to the Higgs
effective potential parameters at the one loop level is given by
the sum of the results for $\mu^{2}$ and $\lambda$ in the fermion
sector, eqs.~(\ref{eq:muparameter}) and~(\ref{eq:lambdaTOP}), and
the corresponding results for the gauge boson contributions in
\emph{Model I}, eqs.~(\ref{eq:muM1}) and~(\ref{eq:lambdaM1}), or
\emph{Model II}, eqs.~(\ref{eq:muM2}) and~(\ref{eq:lambdaM2}).

\section{Numerical results and discussion}

In this section we discuss about the previous results and make
some comments on the constraints that they could  impose on the LH
parameter space in order  to reproduce the SM Higgs potential. It
is well known that this potential has a minimum when
$\mu^2=\lambda v^2$. Furthermore, $\mu$ is forced by data to be at
most of order $200$ GeV. By imposing these conditions we can
obtain the corresponding allowed region of the parameter space of the LH model.
For example, in \cite{ATP} we have obtained that the lowest
allowed value of $\mu$ was of order $500$ GeV considering only the
third generation quark sector. Therefore additional contributions
are required. In this work we have computed the complete one-loop
contributions to the Higgs potential in the framework of the LH
model. However we also found that higher-loops scalar
contributions are still needed in order to get a Higgs mass light
enough to be compatible with the experimental constraints.

If we want to study the allowed region of the parameter space in
these models, we should also take into account other constraints
imposed by requiring the consistency of the LH models with
electroweak precision data. There exist several studies of the
corrections to electroweak precision observables in the Little
Higgs models, exploring whether there are regions of the parameter
space in which the model is consistent with
data~\cite{Schmaltz,review1,Logan,Peskin,Csaki1,Csaki2,
LoganP,EWPO1,recentpheno}.
In the  \emph{Model I} with
a gauge group $SU(2)\times SU(2) \times U(1) \times U(1)$ one has
a multiplet of heavy $SU(2)$ gauge bosons and a heavy $U(1)$ gauge
boson. The last one leads to large electroweak corrections and some
problems with the direct observational bounds on $Z'$ bosons from
Tevatron~\cite{Csaki1,Csaki2}. Then, a very strong bound on
the symmetry breaking scale $f$, $f>4$ TeV at $95 \%$ C.L, is
found~\cite{Csaki1}. This bound is lowered to $1-2$ TeV
for some region of the parameter space~\cite{Csaki2} by gauging only
$SU(2)\times SU(2) \times U(1)$ (\emph{Model II}). In the following, we will
adopt this model and we consider both $f$ about $1$ TeV and $4$ TeV in
the numerical analysis.

For the \emph{Model II} the obtained $\mu$  and $\lambda$
 depend on the heavy top mass $m_{T}$, the heavy gauge-bosons
 mass $M_{W'}$, the coupling constant $\lambda_T$, the
symmetry breaking scale $f$ and the cutoff $\Lambda$. In addition,
one has the mixing angles $\psi$ (for the $SU(2)$ group).
Since we have only one $U(1)$ group, we do not have to consider neither
$M_{B'}$ nor the mixing angle $\psi'$ and $g''$.

These LH model parameters can be bounded as follows:
From the top mass it is possible to set the bounds on the couplings
$\lambda_1, \lambda_2 \geq m_t/v$ or $\lambda_1 \lambda_2 \geq 2
(m_t/v)^2$ ~\cite{Logan}. As a consequence, we get the bound
$\lambda_T \gsim$ 0.5~\cite{ATP}.
On the other hand, in order to avoid a large amount of fine-tuning in
the Higgs potential one has to require  $m_T\lsim$ 2.5
TeV~\cite{Cohen,Peskin}. If $m_T$ is greater than about 2 TeV, the
cancellation of the one-loop quadratic divergences from the top
sector to the Higgs boson mass requires some tuning to give an
answer for $m_H$ below 220 GeV. This cancellation depends on the
relation $m_{T}=f\sqrt{\lambda_{1}^{2}+\lambda_{2}^{2}}$.
Since $m_{T}$ grows linearly with  $f$,
then $f$ should be lesser than about one TeV~\cite{ATP}.
Finally, $\Lambda$ is restricted by the
condition $\Lambda \sim 4 \pi f$~\cite{relationlambdaf}.
Taking into account these
restrictions on the parameters $\lambda_T$, $f$
 and $\Lambda$, we set as first the following ranges:
0.5 $< \lambda_T < 2$, 0.8 TeV $<f<1$ TeV (which implies a heavy
top mass of about 2.5 TeV) and accordingly $10$ TeV $<\Lambda< 12$
TeV. We have checked that these ranges of the parameters are
compatible with the predictions for corrections to the
best-measured observables, the on-shell mass of the W, the
effective mixing angle in $Z^0$ decay asymmetries and the leptonic
width of the $Z^0$, as given in~\cite{Peskin}. In the above paper
the corrections from heavy gauge bosons are included but those
possible corrections coming from a vev of the scalar $SU(2)$
triplet are not considered. A more detailed analysis can be found
in~\cite{Csaki2}.

Then, we also include in our numerical analysis
a discussion on the allowed region of the LH
parameter space for the case of $f=4$ TeV. As established
in~\cite{Csaki1,Csaki2}, this
value of the symmetry breaking scale $f$ is also allowed
by the precision electroweak observables.
Notice that this value of $f$ implies that $m_T$ is
always greater than 5.7 TeV, when $\lambda_T > 0.5$.
A fine-tuning of $0.8\%$ is estimated for a Higgs mass
of $200$ GeV~\cite{Csaki1}. Besides, one gets $M_{W'} > 2.6$ TeV.

\begin{figure}[tb]
\begin{center}
\epsfig{file=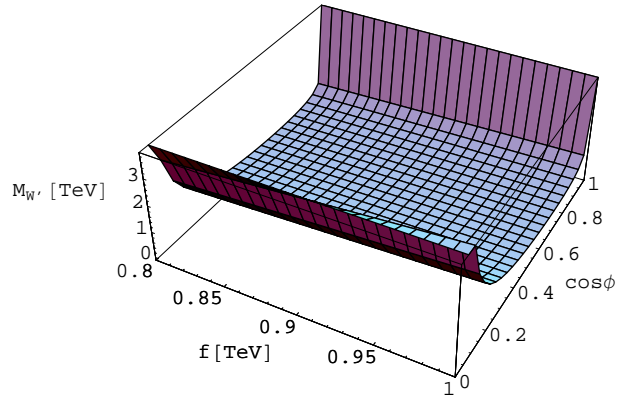,scale=0.8}
\end{center}\vspace*{-0.6cm}
\caption{$M_{W'}$ as a function of $\cos \psi$ and $f$, with
0.8 TeV $< f < 1$ TeV.}
\label{figmasas}
\end{figure}
Let us discuss now briefly how the heavy gauge boson masses depend
on mixing angles $\psi$. We know that $M_{W'}$ is of the order of
$f\sim$ TeV (see eq.(\ref{eq:heavyGB})) and from these equations we can obtain
restrictions on the mixing angles. In Fig.\ref{figmasas} we show
the dependence of $M_{W'}$ on $\cos \psi$ and the scale $f$.
We found that 0.5 $< \cos \psi <$ 0.8 implies masses
smaller than 0.6 TeV and then these values for $\cos \psi$ can be
ruled out.
From this results we get the preferred ranges: $\cos \psi<$ 0.5
or $\cos \psi>$ 0.7.

\begin{figure}[t]
\begin{center}
\epsfig{file=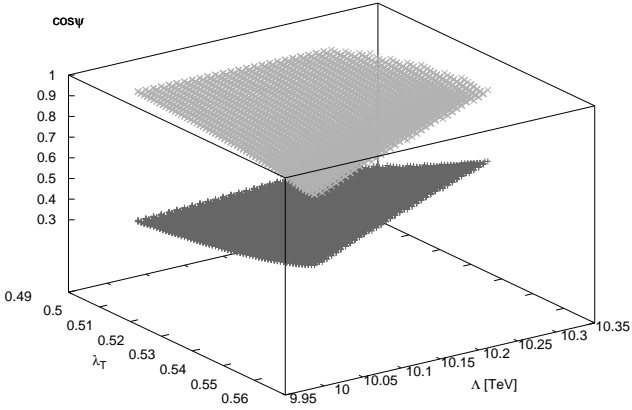,scale=0.37,angle=270}\\
\end{center}\vspace*{-0.3cm}
\caption{Values of $\lambda_{T}$, $\Lambda$ and $\cos\psi$, with 0.5 $<
 \lambda_T < 2$, $10$ TeV $<\Lambda< 12$ TeV, $0<\cos\psi< 1$, $M_{W'}>0.7$ TeV
  and $f= 0.95$ TeV, which satisfy the condition
$\mu^{2}=\lambda v^{2}$.} \label{figM2}
\end{figure}

Taking into account the above bounds on the LH model parameters we
now focus on obtaining the corresponding  $\mu$ values according to
our previous one-loop computation which includes both, the fermionic
and gauge bosons contributions.  We find that, in the case of the
\emph{Model II}, the lowest allowed value for $\mu$ is
$\mu=$0.34 TeV being $\lambda_{T}$=0.7,
$f=$0.8 TeV, $\Lambda=$11.95 TeV and $\cos \psi=$0.2.
However, as discussed in \cite{ATP}, it is also
needed to add in principle the additional constraint
$\mu^{2}=\lambda v^{2}$. Then in Fig.\ref{figM2} we show as an
example the allowed regions for the \emph{Model II}. Two different
regions can be found. This is due to the mixing angle  coming from
the heavy gauge boson mass. In this case the lowest value for
$\mu$ is $\mu=$0.491 TeV corresponding to $\lambda_{T}=$0.55,
$f=$0.95 TeV, $\Lambda$=10 TeV and $\cos \psi=$0.47.
Therefore it is clear that the condition $\mu^{2}=\lambda v^{2}$
is relevant in order to constraint the possible values of the LH
model parameters.

\begin{figure}[t]
\begin{center}
\epsfig{file=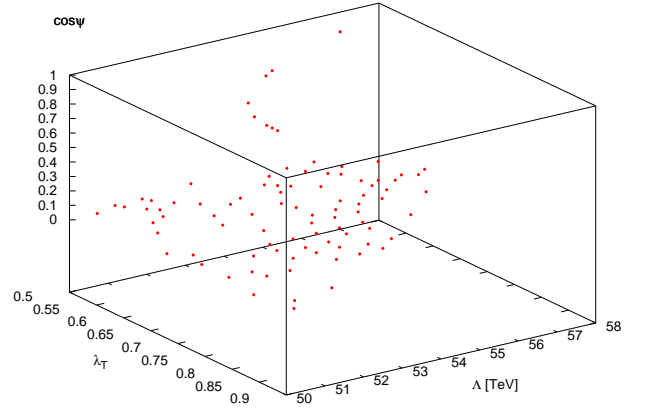,scale=0.37,angle=270}\\
\end{center}\vspace*{-0.3cm}
\caption{Values of $\lambda_{T}$, $\Lambda$ and $\cos\psi$, with 0.5 $<
 \lambda_T < 2$, $0<\cos\psi< 1$ and $f=4$ TeV, which satisfy the condition
$\mu^{2}=\lambda v^{2}$.} \label{figM2f4}
\end{figure}

A similar analysis have been done for $f=4$ TeV.
The results are shown in Fig.\ref{figM2f4}. The allowed region is
smaller in this case. The reason for obtaining just some points
of the parameter space allowed by the condition $\mu^{2}=\lambda v^{2}$
is that $\mu^{2}$ has a logarithmical dependence on the energy scale
$\Lambda$ and a linear dependence on $f$ coming from the new heavy particle
masses, while $\lambda$ depends quadratically on $\Lambda/(4\pi f)$.
Therefore, greater values of $\Lambda$ leads to a disadvantageous region
for $\mu^{2}=\lambda v^{2}$. In this case the lowest value for
$\mu$ is $\mu=$0.916 TeV corresponding to $\lambda_{T}=$0.68,
$f=$4 TeV, $\Lambda$=50 TeV and $\cos \psi=$0.086.

\begin{figure}[tb]
\begin{center}
\epsfig{file=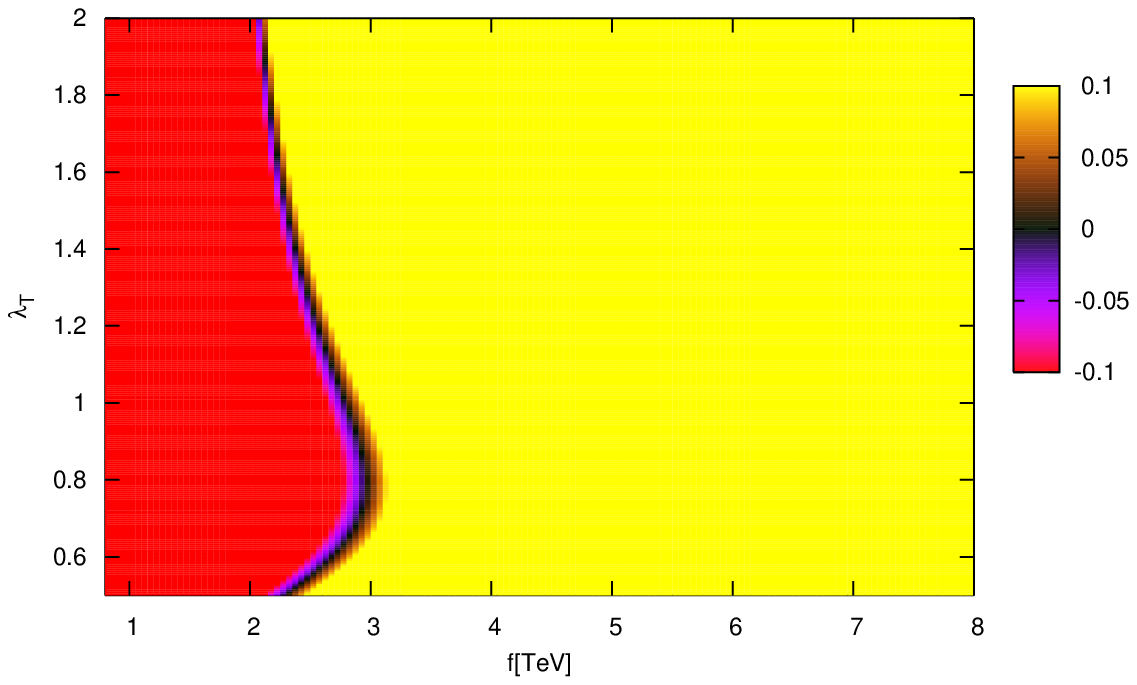,scale=0.65}
\epsfig{file=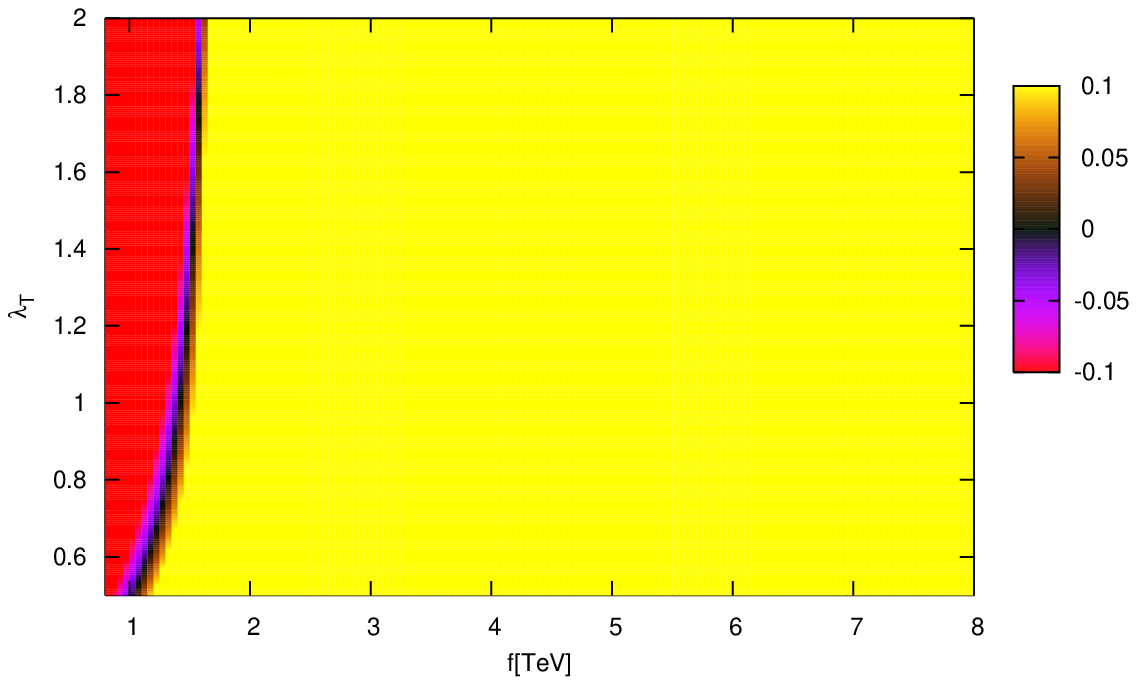,scale=0.65}
\end{center}\vspace*{-0.3cm}
\caption{Contours of the viable regions in the $\lambda_T$-$f$ plane
with the condition $\mu^{2}-\lambda v^{2}=0$. $\cos \psi$ is
fixed to $=0.1$ (top panel) and $0.7$ (bottom panel).} \label{surfaces}
\end{figure}

From the above results, it is clear that is difficult to satisfy
the condition $\mu^{2}=\lambda v^{2}$ with $\mu$ about $200$ GeV,
as expected by the precision electroweak measurements~\cite{LEP}.
We also show in Fig.\ref{surfaces} the contours of the viable
regions in the $\lambda_T$-$f$ plane with the condition
$\mu^{2}=\lambda v^{2}$. The values of the mixing angle $\psi$ are
fixed to the values $\cos \psi=0.1$ (top panel) and $\cos
\psi=0.7$ (bottom panel). We check that the results for $\cos
\psi$ closed to $1$, i.e $\cos \psi=0.995$, are similar to the
ones for $\cos \psi=0.1$. The condition $\Lambda \lesssim 4 \pi f$
is imposed. One can see that values of $f$ around $1-3$ TeV are
the preferred ones for our selected choices of the LH parameters.
However the $\mu$ values are higher than about $350$ GeV for all
cases. Therefore, it is clear that it is not enough to consider
the one-loop effective potential of the
 LH model.

\section{Conclusions}

In this work we have completed the computation of the one-loop
 effective Higgs potential in the context of two versions
(\emph{Model I} and \emph{Model II}) of the LH model. In
particular we have obtained the values of the radiatively
generated $\mu$ and $\lambda$ parameters. Our computation includes
the effect of virtual heavy quarks $t, b$ and $T$, together with
the heavy and electroweak gauge bosons present in the LH model. We
have also clarify the role of the GB when a cutoff is used to
regulate the ultraviolet divergencies. These GB do not contribute
to the Higgs effective potential at the one-loop level but they do
at higher orders. The values of $\mu$ and $\lambda$ that we get
have the right signs and are compatible in  principle with all the
phenomenological constraints set on the LH model parameter space.

However the values found for the $\mu$ parameter are too high to be
compatible with the expected Higgs mass, which should not be larger
than about 200 GeV according to the electroweak precision data. This
problem is even worst if one takes into account the relation
 $\mu^{2}=\lambda v^{2}$ which must hold on the $\mu$
 and $\lambda$ parameters of the effective Higgs potential to reproduce the SM. As
 a conclusion the low-energy,  one-loop  effective
potential of the
 LH model cannot reproduce the SM potential with a low enough Higgs
 mass to agree with the standard expectations. However there
 are some indications  suggesting that higher order
 GB loops could reduce the Higgs boson mass so that complete compatibility
 with the experimental constraints can be obtained. Work is in
 progress in order to check if this is really the case \cite{Future}.

\section{Appendix A}

From (\ref{L0}) it is possible to find the gauge bosons couplings
to doublet Higgs, needed for our computations, which turn to be:

\begin{itemize}
\item Massless gauge bosons-massless gauge bosons:
\begin{eqnarray}
&&\frac{g^{2}}{4}(-\frac{1}{3f^{2}}(HH^{\dag})^{2})W_{\mu}^{a}W^{a\mu}, \hspace{0.25 cm} a=1,2,3,\nonumber \\
\\
&&\frac{g^{'2}}{4}(-\frac{1}{3f^{2}}(HH^{\dag})^{2})B_{\mu}B^{\mu},\\
&-&\frac{1}{2}gg^{'}(H^{0*}H^{+}+H^{0}H^{+*})W_{\mu}^{1}B^{\mu},\\
&-&\frac{i}{2}gg^{'}(H^{0*}H^{+}-H^{0}H^{+*})W_{\mu}^{2}B^{\mu},\\
&-&\frac{1}{2}gg^{'}(H^{0}H^{0*}+H^{+}H^{+*})W_{\mu}^{3}B^{\mu}.
\end{eqnarray}

\item Heavy gauge bosons-massless gauge bosons:
\begin{eqnarray}
&&g_{W'W}HH^{\dag} W_{\mu}^{'a}W^{a\mu}, \hspace{0.25 cm}\\
&&g_{B'B}HH^{\dag} B_{\mu}^{'}B^{\mu},\\
&-&g_{W'B}(H^{0*}H^{+}+H^{0}H^{+*})W_{\mu}^{'1}B^{\mu},\\
&-i&g_{W'B}(H^{0*}H^{+}-H^{0}H^{+*})W_{\mu}^{'2}B^{\mu},\\
&-&g_{W'B}(H^{0}H^{0*}-H^{+}H^{+*})W_{\mu}^{'3}B^{\mu},\\
&-&g_{B'W}(H^{0*}H^{+}+H^{0}H^{+*})B'_{\mu}W^{\mu 1},\\
&-i&g_{B'W}(H^{0*}H^{+}-H^{0}H^{+*})B'_{\mu}W^{\mu 2},\\
&-&g_{B'W}(H^{0}H^{0*}-H^{+}H^{+*})B'_{\mu}W^{\mu 3}.
\end{eqnarray}

\begin{widetext}
\item Heavy gauge bosons-heavy gauge bosons:
\begin{flushleft}
\begin{eqnarray}
&&\left\{g_{W'W'}\left[3(g_{1}^{2}+g_{2}^{2})^{2}(H^{0*}H^{+}+H^{0}H^{+*})^{2}
+4g_{1}^{2}g_{2}^{2}(HH^{\dag})^{2}\right]-\frac{1}{4}g^{2}HH^{\dag}\right\}W_{\mu}^{'1}W^{'1\mu},\\
&&\left\{g_{W'W'}\left[-3(g_{1}^{2}+g_{2}^{2})^{2}(H^{0*}H^{+}-H^{0}H^{+*})^{2}
+4g_{1}^{2}g_{2}^{2}(HH^{\dag})^{2}\right]-\frac{1}{4}g^{2}HH^{\dag}\right\}W_{\mu}^{'2}W^{'2\mu},\\
&&\left\{g_{W'W'}\left[-12(g_{1}^{2}+g_{2}^{2})^{2}(H^{0}H^{0*}H^{+}H^{+*})
+(10g_{1}^{2}g_{2}^{2}+3(g_{1}^{4}+g_{2}^{4}))(HH^{\dag})^{2}\right]-\frac{1}{4}g^{2}
HH^{\dag}\right\}W_{\mu}^{'3}W^{'3\mu},\\
&&\left\{g_{B'B'}\left[3(g_{1}^{'4}+g_{2}^{'4})+10g_{1}^{'2}g_{2}^{'2}
\right](HH^{\dag})^{2}-\frac{1}{4}g^{'2}HH^{\dag}\right\}B^{'}_{\mu}B^{'\mu},\\
&&g_{W'B'}(H^{0*}H^{+}+H^{0}H^{+*})W_{\mu}^{'1}B^{'\mu},\\
&i&g_{W'B'}(H^{0*}H^{+}-H^{0}H^{+*})W_{\mu}^{'2}B^{'\mu}, \\
&&g_{W'B'}(H^{0}H^{0*}+H^{+}H^{+*})W_{\mu}^{'3}B^{'\mu}.
\end{eqnarray}
\end{flushleft}
\end{widetext}
\end{itemize}
The different couplings appearing above are given by:

$g_{W'W'}=\frac{1}{48 f^{2}(g_{1}^{2}+g_{2}^{2})},$ \hspace{1cm}
$g_{B'B'}=\frac{1}{48 f^{2}(g_{1}^{'2}+g_{2}^{'2})},$

$g_{W'W}=\frac{g(g_{1}^{2}-g_{2}^{2})}{4\sqrt{g_{1}^{2}+g_{2}^{2}}},$ \hspace{1.5cm}
$g_{B'B}=\frac{g'(g_{1}^{'2}-g_{2}^{'2})}{4\sqrt{g_{1}^{'2}+g_{2}^{'2}}},$

$g_{W'B}=\frac{g'(g_{1}^{2}-g_{2}^{2})}{4\sqrt{g_{1}^{2}+g_{2}^{2}}},$ \hspace{1.5cm}
$g_{B'W}=\frac{g(g_{1}^{'2}-g_{2}^{'2})}{4\sqrt{g_{1}^{'2}+g_{2}^{'2}}}.$

$g_{W'B'}=\frac{g_{1}^{'2}g_{2}^{2}+g_{1}^{2}g_{2}^{'2}}{4\sqrt{(g_{1}^{2}+g_{2}^{2})(g_{1}^{'2}+g_{2}^{'2})}},$
\vspace{0.75cm}

\section{Appendix B}
The integrals appearing in our computations are:
\begin{flushleft}
\begin{eqnarray}
I_{0}(M^{2}) &\equiv& \int d\tilde{p} \frac{i}{p^{2}-M^{2}}
  \nonumber   \\
  I_{1}(M^{2}) &\equiv& \int d\tilde{p}
\frac{i}{(p^{2}-M^{2})^{2}}   \nonumber   \\
 I_{2}(M^{2})
&\equiv& \int d\tilde{p} \frac{i}{p^{2}(p^{2}-M^{2})}
\nonumber \\
I_{2}(0) &\equiv& \int d\tilde{p} \frac{i}{p^{4}}  \nonumber   \\
 I_{3}(M_{a}^{2},M_{b}^{2}) &\equiv& \int d\tilde{p}
\frac{i}{(p^{2}-M_{a}^{2})(p^{2}-M_{b}^{2})}. \nonumber
\end{eqnarray}
\end{flushleft}
Using an ultraviolet cutoff $\Lambda$ these integrals are found to
be:
\begin{flushleft}
\begin{eqnarray}
I_{0}(M^{2})&=&\frac{1}{(4\pi)^{2}}\left[\Lambda^{2}-M^{2}\log
\left(1+\frac{\Lambda^{2}}{M^{2}}\right)\right]   \nonumber\\
I_{1}(M^{2})&=&-\frac{1}{(4\pi)^{2}}\left[\log \left(1+
\frac{\Lambda^{2}}{M^{2}}\right)-\frac{1}{1+\frac{M^{2}}{\Lambda^{2}}}\right]
\nonumber   \\
 I_{2}(M^{2})&=&-\frac{1}{(4\pi)^{2}}\log
\left(1+\frac{\Lambda^{2}}{M^{2}}\right)   \nonumber   \\
I_{2}(0)&=&-\frac{1}{(4\pi)^{2}}\log
\left(\frac{\Lambda^{2}}{m^{2}}\right)   \nonumber   \\
I_{3}(M_{a}^{2},M_{b}^{2})&=&-\frac{1}{(4\pi)^{2}}\frac{1}{M_{a}^{2}-M_{b}^{2}}
\left[M_{a}^{2}\log
\left(1+\frac{\Lambda^{2}}{M_{a}^{2}}\right)\right. \nonumber
\\ &-& \left. M_{b}^{2}\log
\left(1+\frac{\Lambda^{2}}{M_{b}^{2}}\right)\right]  \nonumber
\end{eqnarray}
\end{flushleft}
where $m$ is an infrared cutoff.

\vspace{.1cm}

 {\bf Acknowledgements:}
This work is supported by DGICYT (Spain) under project number
BPA2005-02327. The work of S.P. is supported by the {\it
I3P}-Contract of CSIC at IFIC - Instituto de F\'{\i}sica
Corpuscular, Valencia. The work of L. Tabares-Cheluci is supported
by FPU grant from the Spanish M.E.C.

\end{document}